\documentclass[final]{phyproc}
\usepackage{graphicx}
\newcommand{\bea}{\begin{eqnarray}}
\newcommand{\eea}{\end{eqnarray}}
\usepackage{amsmath}

\def\beq{\begin{equation}} 
\def\eeq{\end{equation}} 
\def\to{\rightarrow}

\def\b0{\beta_0}

\def\beeq{\begin{eqnarray}}
\def\eeeq{\end{eqnarray}}

\def\ca{{\cal A}}
\def\sh{\hat{s}}
\def\Re{{\rm Re}}
\def\Im{{\rm Im}}

\layoutstyle{8x11single}

\begin{document}

\title{QCD Phenomenology}


\author{Roger Jos\'e Hern\'andez-Pinto}{
  address={
  Departamento de F\'isica, FCEyN, Universidad de Buenos Aires, \\
  (1428) Pabell\'on 1 Ciudad Universitaria, Capital Federal, Argentina and \\
 Instituto de F\'isica Corpuscular, Universitat de Val\`encia - Consejo Superior de Investigaciones Cient\'ificas, 
 Parc Cient\'ific, E-46980 Paterna, Valencia, Spain
 }}


\begin{abstract}
Quantum Chromodynamics is the most successful theory in particle physics. The understanding of all different signals at hadron colliders have been achieved due to the correct interpretation of the theory. In this paper we review some basic features of the theory of strong interactions and how it could be used in order to provide phenomenological distributions for the Large Hadron Collider. The main results presented in here can be found in Ref~\cite{deFlorianFid:2013xc}.
\end{abstract}

\maketitle

\section{Introduction}
\label{sec:Introduction}
The Large Hadron Collider (LHC) is the biggest machine ever built by humanity. Its main purpose, the discovery of the last remaining particle of the Standard Model (SM) was achieved in 2012.  Theorists and experimentalists from all around the globe made an enormous effort in order to succeed on this task. Experiments are achieving a very high precision in all measurements, and they are now also pushing theorists to provide phenomenological SM predictions at the same level of accuracy. Besides that, the new era of the LHC is coming, and the disentanglement of the properties of the Higgs particle or the discovery of new physics, require theoretical Monte Carlo simulations, of signal and background, at the highest possible precision.

The SM of particle physics is based on a gauge theory of $SU(3)_c\times SU(2)_L\times U(1)_Y$ and it has been by far the most precise theory of nature. In particular, the sector of the theory which governs the physics of the LHC is the one related with the $SU(3)_c$. The LHC collides protons at center of mass energies of the order of TeVs. Protons are made of quarks and gluons, elementary particles of the SM. The description of these partons is well understood in the framework of Quantum Chromodynamics (QCD). Unfortunately, QCD cannot be solved completely, and usually in order to make theoretical predictions for hadron colliders, one takes the perturbative version of QCD (pQCD). In the perturbative regime, the coupling associated to $SU(3)_c$ is considered small and the series expansion is allowed. However, at the LHC, this assumption is only valid just in the moment when the collision occurs and when particles are flowing into detectors this statement could not be longer true. In order to compute observables, it is important to know how to include the perturbative part and the non perturbative one in the calculation. In fact, one way to describe the production of a hadron $H$ at the LHC is a mixing between this two sectors,
\begin{eqnarray}
E_H\frac{d^3\sigma}{dp^3_H}=\sum_{a,b,c}f_a\otimes f_b\otimes d\hat\sigma^c_{ab}\otimes D_c^H
\end{eqnarray}
where the sum runs over all partonic channels, $a+b\rightarrow c+X$, with $d\hat\sigma^c_{ab}$ the associated partonic cross section; $f_a$, $f_b$ are the Parton Distribution Functions (PDF) and $D_c^H$ are the Fragmentation Functions (FF). The partonic cross section can be expanded as a power series in the strong coupling $\alpha_S$, and the PDF and the FF are the non-perturbative objects that are extracted directly from the experimental data. 

In order to obtain a phenomenological prediction for the LHC, one needs to convolute the perturbative and non-perturbative part consistently and evolve all parameters to the scale when the process occurs. In this short review, we sketch how it is possible to use pQCD for describing small deviations of SM signals, such as the diphoton channel for the Higgs decay.

\section{Theory}
\label{sec:theory}
The main channel for the Higgs production at hadron colliders is the gluon-gluon fusion, which is mediated principally by a heavy-quark (top-quark) loop~\cite{Georgi:1977gs}. A lot of work have been done around this channel due to the great relevance for the LHC including up to next-to-next-to-next-to-leading (N$^3$LO) QCD corrections~\cite{Dawson:1990zj,Djouadi:1991tk,Spira:1995rr,Harlander:2002wh,Anastasiou:2002yz,Ravindran:2003um,Anastasiou:2005qj,Anastasiou:2007mz,Catani:2007vq,Grazzini:2008tf,Mazzitelli:2014}.
State of the art computations for this channel \cite{deFlorian:2012yg} include electroweak corrections at NLO \cite{ew,Actis:2008ug} and soft gluon resummation to next-to-next-to leading logarithmic accuracy \cite{Catani:2003zt}.
In this note, we will focus on the diphoton decay channel of the Higgs. This channel is very interesting because its branching fraction is extremely small but it has a clean experimental signal then, it was one of the most significant channels for the Higgs analysis. 

A phenomenological calculation for the LHC is not completed without the knowledge of the background, thus the corresponding background for diphoton production was also computed up to NNLO accuracy\cite{Catani:2011qz}. Small deviations could appear when the analysis is achieved to higher precision and also when the interference between signal and background is considered. In the following we focus on calculating the interference between signal and background for the diphoton channel in proton proton collisions.

The interference of the resonant process $ij \to X+H  \to \gamma \gamma $ with the continuum QCD background $ij \to X+\gamma\gamma $ induced by quark loops can be expressed at the level of the partonic cross section as:
\begin{eqnarray}
\delta\hat{\sigma}_{ij\to X+ H\to \gamma\gamma} &=& 
-2 (\sh-m_H^2) { \Re \left( \ca_{ij\to X+H} \ca_{H\to\gamma\gamma} 
                          \ca_{\rm cont}^* \right) 
        \over (\sh - m_H^2)^2 + m_H^2 \Gamma_H^2 }
\nonumber\\
&& 
-2 m_H \Gamma_H { \Im \left( \ca_{ij\to X+H} \ca_{H\to\gamma\gamma} 
                          \ca_{\rm cont}^* \right)
        \over (\sh - m_H^2)^2 + m_H^2 \Gamma_H^2 } \,,
\label{intpartonic}
\end{eqnarray}
where $\sh$ is the partonic invariant mass, $m_H$ and $\Gamma_H$ are the Higgs mass and decay width respectively. 

It has been noticed in \cite{Dixon:2003yb,Dicus:1987fk}, that the real part of the amplitudes is odd in $\sh$ around $m_H$, therefore its effect on the total $\gamma\gamma$ rate is subdominant. On the other hand, for the gluon-gluon partonic subprocess~\cite{Dicus:1987fk} it was found that the imaginary part of the corresponding one-loop amplitude has a quark mass suppression for the relevant helicity combinations. 

LHC collides protons which contain gluons and quarks then, the full diphoton channel has to be considered theoretically as a sum of all possible initial states: gluon-gluon ($gg$), quark-gluon ($qg$) and quark-antiquark ($q\bar{q}$). Even if the contribution will be subdominant for the $qg$ and $q\bar{q}$ due to PDF, these channels could bring sensible discrepancies in the full analysis, therefore it is worth studying completely the diphoton production at the order $\mathcal{O}(\alpha_S^2)$.

\section{Results}
\label{sec:resul}
\begin{figure}[!h]
\includegraphics[scale=0.55]{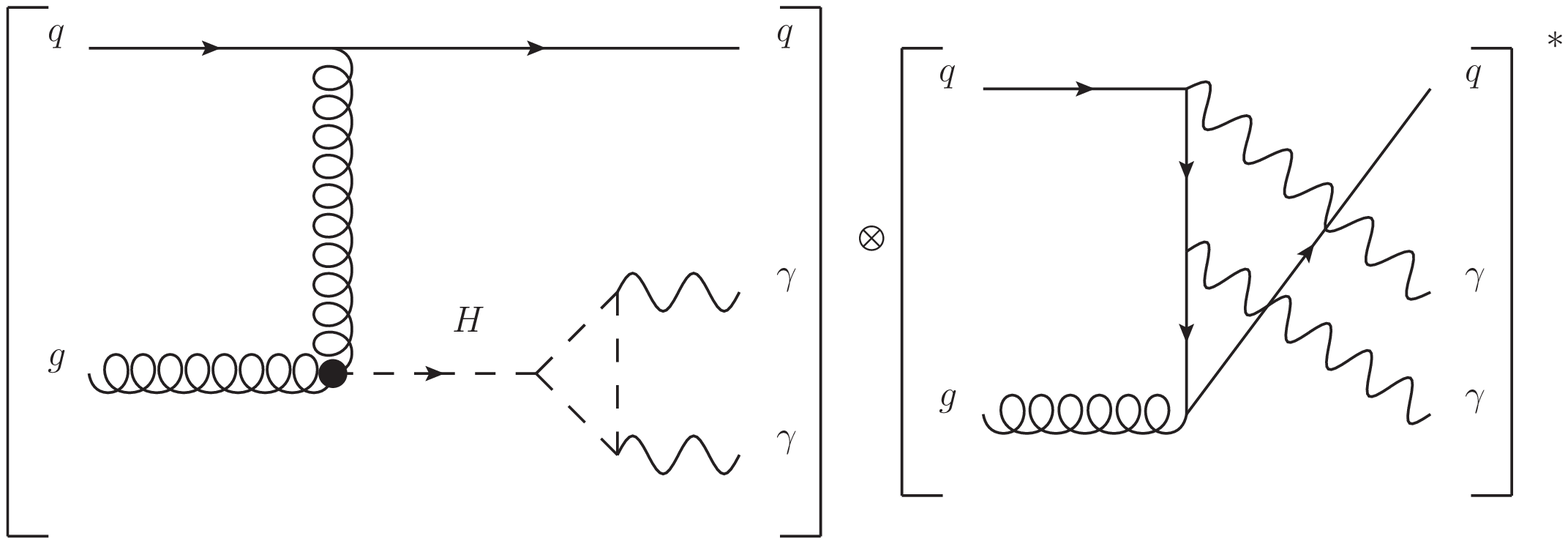}
\caption{ Sample of Feynman diagrams contributing to the interference}
\label{fig:qgall}
\end{figure}

The starting point consists on computing all amplitudes using the Mathematica package ${ FeynArts}$ \cite{Hahn:2000kx} for all interferences, $gg$, $qg$ and $q\bar{q}$. Then, combining with the package ${ FeynCalc}$  \cite{Mertig:1990an}, it is possible to obtain the final squared matrix elements. For simplicity, the production amplitudes are computed within the effective Lagrangian approach for the $ggH$ coupling (relying in the infinite top mass limit), approximation known to work at the few percent level for the process of interest. A sample of the Feynman diagrams for the $qg$ interference channel are shown in Figure \ref{fig:qgall}. 

One might notice that in the $qg$ channel there is an extra parton in the final state. In order to find the contribution to the cross section of the diphoton channel, this parton has to be integrated out. Infrared singularities appear on the amplitudes individually, however the interference remains finite after the phase space integration and its contribution is truly at tree-level.

Having the matrix squared amplitudes, we performed a convolution of the partonic cross-section with the PDF. We used the MSTW2008 LO set \cite{Martin:2009iq}  (in the five massless flavours scheme), and the one-loop expression of the strong coupling constant, setting the factorization and renormalization scales to the diphoton invariant mass $\mu_F=\mu_R= M_{\gamma\gamma}$. The decay into two photons is treated exactly and we set $\alpha=1/137$.
For the Higgs boson we use $m_H=125\,\text{GeV}$ and $\Gamma_H=4.2\,\text{MeV}$. For all the histograms we implemented an asymmetric cut on the transverse momentum of the photons: $p_{T, \gamma}^{hard (soft)} \geq 40(30)  \text{ GeV}$. Their pseudorapidity was also constrained to be in the region of $|\eta_\gamma | \leq 2.5$.
We also applied the standard isolation prescription for the photons,  requesting that the transverse hadronic energy deposited within a cone of size $R=\sqrt{\Delta \phi^2+\Delta \eta^2}<0.4$ around the photon should satisfy $p_{T,had} \leq 3 \text{ GeV}$. In addition, we reject all the events with $R_{\gamma\gamma} < 0.4$. 

\begin{figure}[!h]
\includegraphics[scale=1.4]{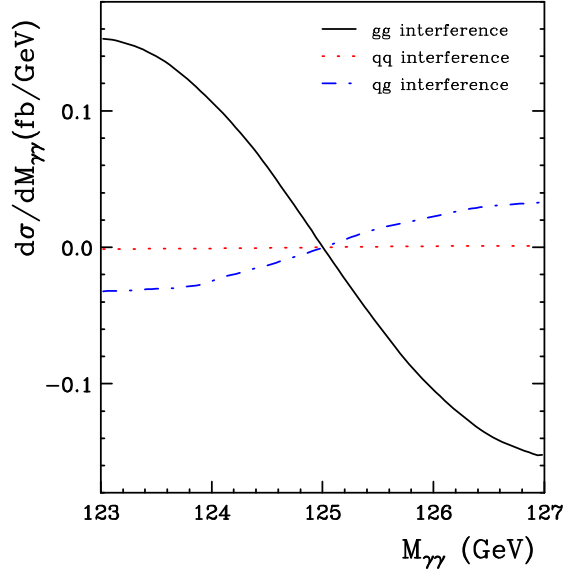}
\caption{Diphoton invariant mass distribution for the interference terms including the smearing effects which simulate the detector (Gaussian function of width $1.7 \, \text{GeV}$). The solid line represents the $gg$ channel contribution, the dashed line represents the $qg$ channel, and the dotted one, the $q \bar q$.}
\label{fig:int_gauss}
\end{figure}

In order to simulate the response of the detector, we convoluted the partonic cross-section with a Gaussian function of mass resolution width $\sigma_{\text{MR}}=1.7\,\text{GeV}$ following the procedure showed in Ref.~\cite{Martin:2012xc}. The corresponding results are presented in Figure \ref{fig:int_gauss}. We can observe that the magnitude of the interference is reduced, but the position of the peak (and dip) is moved as much as $2  \text{ GeV}$. 
From Figure~\ref{fig:int_gauss} one can conclude that the displacement of the invariant mass peak will be driven by: the width of the gaussian, the magnitude and sign of the interference.

\begin{figure}[!h] 
\includegraphics[scale=1.4]{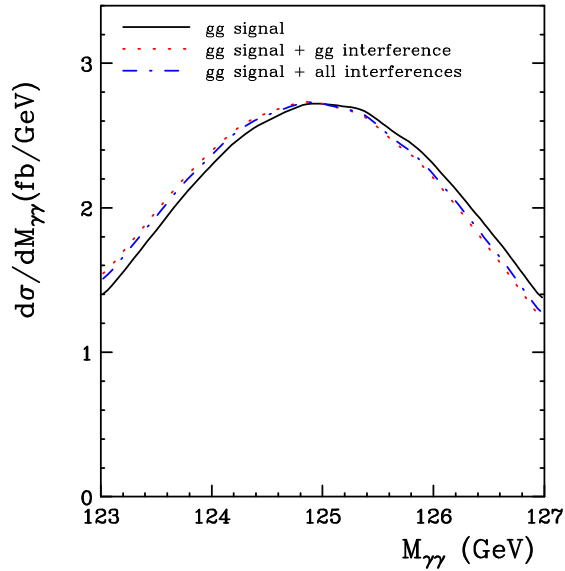}
\caption{Diphoton invariant mass distribution including the smearing effects of the detector (Gaussian function of width $1.7 \, \text{GeV}$). The solid line corresponds to the signal-only contribution. The dotted line corresponds to the distribution  after adding the $gg$ interference term, and the dashed line represents the complete Higgs signal plus all three interference contributions ($gg$, $qg$ and $q \bar q$).}
\label{fig:shift}
\end{figure}

We present in Figure \ref{fig:shift} the corresponding results after adding the Higgs signal. The solid curve corresponds to the signal cross-section, without the interference terms, but including the detector smearing effects. As expected, the (signal) Higgs peak remains at $M_{\gamma\gamma}=125 \, \text{GeV}$. When adding the $gg$ interference term, we observe a shift on the position of the peak in about $150 \, \text{MeV}$ towards a lower mass (dotted), as found in Ref.~\cite{Martin:2012xc}. If we also add the $qg$ and $q \bar q$ contributions (dashed), the peak is shifted slightly (of the order of just a few MeV) back towards a higher mass region because of the opposite sign of the amplitudes. 

To have a rough estimation of the theoretical uncertanties, we varied the factorization and renormalization scales between $\frac{1}{2} M_{\gamma\gamma}$ and $\frac{1}{2} M_{\gamma\gamma}$. 
There are large variations, of about $80 \%$, in all the channels. 
Given the fact that very large K-factors are observed in both the signal and the background, one might expect a considerable increase in the interference as well.


\section{Conclusions}
\label{sec:conc}
In this note, we have presented an example of how pQCD could help to analyse the small corrections to theoretical predictions of experimental observables. We showed the shift in the invariant mass distribution for the diphoton channel at the LHC, when the interference between the signal and background is considered in a full QCD calculation at the order $\mathcal{O}(\alpha_S^2)$.


\begin{theacknowledgments} 
	I would like to extend my acknowlendment to all members of the University of Buenos Aires who I shared the opportunity to collaborate on this work.  This work was supported by CONACyT-Mexico, the Research Executive Agency (REA) of the European Union under the Grant Agreement number PITN-GA-2010-264564 (LHCPhenoNet), the Spanish Government and EU ERDF funds (FPA2011-23778 and CSD2007-00042 CPAN).

\end{theacknowledgments}

\bibliographystyle{phyproc} 

\begin{thebibliography}{9}


\bibitem{deFlorianFid:2013xc}
  D.~de~Florian, N.~Fidanza, R.~J.~Hernández-Pinto, J.~Mazzitelli, Y.~Rotstein-Habarnau, G.~F.~R.~Sborlini 
  Eur.Phys.J. {\bf C73} (2013) 2387 
  [arXiv:1303.1397 [hep-ph]].


\bibitem{Georgi:1977gs}
  H.~M.~Georgi, S.~L.~Glashow, M.~E.~Machacek and D.~V.~Nanopoulos,
  Phys.\ Rev.\ Lett.\  {\bf 40} (1978) 692.

\bibitem{Dawson:1990zj}
  S.~Dawson,
  Nucl.\ Phys.\  B {\bf 359} (1991) 283.

\bibitem{Djouadi:1991tk}
A.~Djouadi, M.~Spira and P.~M.~Zerwas,
Phys.\ Lett.\ B {\bf 264} (1991) 440.



\bibitem{Spira:1995rr}
  M.~Spira, A.~Djouadi, D.~Graudenz and P.~M.~Zerwas,
  Nucl.\ Phys.\  B {\bf 453} (1995) 17.


\bibitem{Harlander:2002wh}
  R.~V.~Harlander and W.~B.~Kilgore,
  Phys.\ Rev.\ Lett.\  {\bf 88} (2002) 201801.

\bibitem{Anastasiou:2002yz}
  C.~Anastasiou and K.~Melnikov,
  Nucl.\ Phys.\  B {\bf 646} (2002) 220.

\bibitem{Ravindran:2003um}
  V.~Ravindran, J.~Smith and W.~L.~van Neerven,
  Nucl.\ Phys.\  B {\bf 665} (2003) 325.



\bibitem{Anastasiou:2005qj}
  C.~Anastasiou, K.~Melnikov and F.~Petriello,
  Phys.\ Rev.\ Lett.\  {\bf 93} (2004) 262002,
  Nucl.\ Phys.\  B {\bf 724} (2005) 197.

\bibitem{Anastasiou:2007mz}
  C.~Anastasiou, G.~Dissertori and F.~Stockli,
  JHEP {\bf 0709} (2007) 018.


\bibitem{Catani:2007vq}
  S.~Catani and M.~Grazzini,
  Phys.\ Rev.\ Lett.\  {\bf 98} (2007) 222002.

\bibitem{Grazzini:2008tf}
  M.~Grazzini,
  JHEP {\bf 0802} (2008) 043.

\bibitem{Mazzitelli:2014}
D.~de Florian, J.~Mazzitelli, S.~Moch and V.~Vogt,
  [arXiv:1408.6277 [hep-ph]].

\bibitem{deFlorian:2012yg}
  D.~de Florian and M.~Grazzini,
  Phys.\ Lett.\ B {\bf 718} (2012) 117
  [arXiv:1206.4133 [hep-ph]].



\bibitem{ew}
U.~Aglietti, R.~Bonciani, G.~Degrassi and A.~Vicini, 
Phys.\ Lett. B {\bf 595} (2004) 432;
G.~Degrassi and F.~Maltoni,
Phys.\ Lett. B {\bf 600} (2004) 255;
U.~Aglietti, R.~Bonciani, G.~Degrassi and A.~Vicini,
contributed to the {\em TeV4LHC Workshop}, Brookhaven, Upton, New York, february 2005, arXiv:hep-ph/0610033.


\bibitem{Actis:2008ug}
S.~Actis, G.~Passarino, C.~Sturm and S.~Uccirati,
Phys.\ Lett.\  B {\bf 670} (2008) 12;
 Nucl.\ Phys.\  B {\bf 811} (2009) 182.
\bibitem{Catani:2003zt}
S.~Catani, D.~de Florian, M.~Grazzini and P.~Nason,
JHEP {\bf 0307} (2003) 028.


\bibitem{Catani:2011qz}
  S.~Catani, L.~Cieri, D.~de Florian, G.~Ferrera and M.~Grazzini,
  Phys.\ Rev.\ Lett.\  {\bf 108} (2012) 072001
  [arXiv:1110.2375 [hep-ph]].
  
  



\bibitem{Dixon:2003yb}
  L.~J.~Dixon and M.~S.~Siu,
  Phys.\ Rev.\ Lett.\  {\bf 90} (2003) 252001
  [hep-ph/0302233].

\bibitem{Dicus:1987fk}
  D.~A.~Dicus and S.~S.~D.~Willenbrock,
  Phys.\ Rev.\ D {\bf 37} (1988) 1801.


\bibitem{Hahn:2000kx}
  T.~Hahn,
  Comput.\ Phys.\ Commun.\  {\bf 140} (2001) 418
  [hep-ph/0012260].

\bibitem{Mertig:1990an}
  R.~Mertig, M.~Bohm and A.~Denner,
  Comput.\ Phys.\ Commun.\  {\bf 64} (1991) 345.




\bibitem{Martin:2009iq}
  A.~D.~Martin, W.~J.~Stirling, R.~S.~Thorne and G.~Watt,
  Eur.\ Phys.\ J.\ C {\bf 63} (2009) 189
  [arXiv:0901.0002 [hep-ph]].


\bibitem{Martin:2012xc}
  S.~P.~Martin,
  Phys.\ Rev.\ D {\bf 86} (2012) 073016
  [arXiv:1208.1533 [hep-ph]].






  














  

\end{thebibliography}

\end{document}